\documentstyle[12pt]{article}
\makeatletter
\newif\ifabstract \abstractfalse
\newif\ifpreprint \preprintfalse
\def\preprint{\preprinttrue
\topmargin 0 pt     
\textheight 42\baselineskip	
\advance\textheight by \topskip
\oddsidemargin -10 pt      
\evensidemargin 10 pt      
\marginparwidth 1 in       
\oddsidemargin 0 in      
\evensidemargin 0 in
\marginparwidth 0.75 in
\textwidth 16.5 truecm 
\pagestyle{plain}
}
\def\final{\preprintfalse
\topmargin 0 pt     
\textheight 46\baselineskip	
\advance\textheight by \topskip
\oddsidemargin -10 pt      
\evensidemargin 10 pt      
\marginparwidth 1 in       
\oddsidemargin 0 in      
\evensidemargin 0 in
\marginparwidth 0.75 in
\textwidth 16.5 truecm 
\headsep\z@ \headheight\z@
\pagestyle{empty}
}
\def\section{\@startsection {section}{1}{\z@}{-3.5ex plus -1ex minus
 -.2ex}{2.3ex plus .2ex}{\bf}*}
\def\subsection{\@startsection{subsection}{2}{\z@}{-3.25ex plus -1ex minus
 -.2ex}{1.5ex plus .2ex}{\sc}*}
\def\paragraph{\@startsection {paragraph}{4}{\z@ }{3.25ex plus 1ex
minus .2ex}{-1em}{\normalsize\sc }}
\def\bibliography#1{\if@filesw\immediate\write\@auxout
  {\string\bibstyle{npb}}\fi
  \if@filesw\immediate\write\@auxout{\string\bibdata{#1}}\fi
  \@input{\jobname.bbl}}
\def\thebibliography#1{\section{References\@mkboth
 {REFERENCES}{REFERENCES}}\list
 {[\arabic{enumi}]}{\settowidth\labelwidth{[#1]}\leftmargin\labelwidth
 \advance\leftmargin\labelsep \itemsep=0pt
 \usecounter{enumi}}
\ifpreprint\else\small\baselineskip=12pt\fi
 \def\newblock{\hskip .11em plus .33em minus .07em}
 \sloppy\clubpenalty4000\widowpenalty4000
 \sfcode`\.=1000\relax}
\def\@citex[#1]#2{%
\if@filesw \immediate \write \@auxout {\string \citation {#2}}\fi
\@tempcntb\m@ne \let\@h@ld\relax \def\@citea{}%
\@cite{%
  \@for \@citeb:=#2\do {%
    \@ifundefined {b@\@citeb}%
      {\@h@ld\@citea\@tempcntb\m@ne{\bf ?}%
      \@warning {Citation `\@citeb ' on page \thepage \space undefined}}%
      {\@tempcnta\@tempcntb \advance\@tempcnta\@ne%
      \@tempcntb\number\csname b@\@citeb \endcsname \relax%
      \ifnum\@tempcnta=\@tempcntb 
        \ifx\@h@ld\relax%
          \edef \@h@ld{\@citea\csname b@\@citeb\endcsname}%
        \else%
          \edef\@h@ld{\ifmmode{-}\else--\fi\csname b@\@citeb\endcsname}%
        \fi%
      \else
        \@h@ld\@citea\csname b@\@citeb \endcsname%
        \let\@h@ld\relax%
      \fi}%
    \def\@citea{,\penalty\@highpenalty\,}%
  }\@h@ld%
}{#1}}
\gdef\@publabel{\hfil}
\gdef\@pubdate{\null}
\gdef\@pubnumber{\null}
\gdef\@author{\null}
\gdef\@title{\null}
\gdef\@abstract{\null}
\long\def\pubdate#1{\gdef\@pubdate{#1}}
\long\def\pubnumber#1{\gdef\@pubnumber{#1}}
\long\def\publabel#1{\gdef\@publabel{#1}}
\long\def\author#1{\gdef\@author{#1}}
\long\def\title#1{\gdef\@title{#1}}
\long\def\abstract#1{\abstracttrue\gdef\@abstract{#1}}
\def\titlerelax{
\let\maketitle\relax
\let\settitleparameters\relax
\let\inittitlepage\relax
\let\finishtitlepage\relax
\let\titlepagecontents\relax
\let\multithanks\relax
\let\titlebaselines\relax
\let\@makepub\relax
\let\@maketitle\relax
\let\@makeauthor\relax
\let\@makeabstract\relax
\let\@maketitlenote\relax
\let\thanks\relax
\let\titlerelax\relax}
\def\titleclean
{\gdef\@titlenote{}
\gdef\@abstract{}
\gdef\@author{}
\gdef\@title{}
\gdef\@pubdate{}\gdef\@pubnumber{}\gdef\@publabel{}
\gdef\@publabel{}
}
\def\maketitle{
{\setlength{\parskip}{0pt}
\nointerlineskip
\let\footnotesize\small
\ifpreprint
\begin{flushright}\@pubnumber \end{flushright}
\fi
\begin{center} \Large \@title \par \end{center}
{\def\and{\smallskip {\normalsize \rm and\smallskip}}
\long\def\address##1{{\def\and{\\\medskip{\rm and}\\\medskip}\medskip
				{\it \\##1\\}}}
{\centering\small \@author} \par}
\small\quotation \@abstract \endquotation \@thanks
}
	\titlerelax \titleclean
	\setcounter{footnote}{0}
}

\preprint
\def\cF{{\cal F}}

\let\c=\chi
\let\d=\delta

\let\g=\gamma

\let\p=\phi

\let\s=\sigma

\let\z=\zeta
\def\@ifhat#1#2{\@ifnextchar ^{\def\@tempa^{#1}\@tempa}{#2}}
\def\conj#1{\@ifhat{\@conj{#1}}{#1^\vee}}
\def\cconj#1{\@ifhat{\@cconj{#1}}{#1^{\vee\vee}}}
\def\@conj#1#2{#1^{#2\vee}}
\def\@cconj#1#2{#1^{#2\vee\vee}}

\def\su{\mathop{\it su}\nolimits}

\def\id{{\bf 1}}

\def\rank{\mathop{\rm rank}\nolimits}

\def\WA{\mathop{\it WA}\nolimits}
\def\WB{\mathop{\it WB}\nolimits}

\def\WC{\mathop{\it WC}\nolimits}
\def\WD{\mathop{\it WD}\nolimits}
\def\WE{\mathop{\it WE}\nolimits}

\def\WG{\mathop{\it WG}\nolimits}
\def\rllap#1{\hbox to 0pt{\hss#1\hss}}

\def\half#1{{\textstyle{#1\over2}}}

\def\nord:#1:{\mathopen{\scriptstyle{\circ \atop \circ}}#1
\mathclose{\scriptstyle{\circ \atop \circ}}}
\def\bra#1.{{\langle#1|}}
\def\ket#1.{{|#1\rangle}}

\def\cca(#1,#2){{\textstyle{#1 \atopwithdelims[] #2}}}
\def\ccb(#1,#2){{\textstyle{#1 \atopwithdelims\{\} #2}}}
\def\ip(#1,#2){{\langle#1|#2\rangle}}
\def\ev(#1,#2){{\left\langle#1,#2\right\rangle}}
\def\cfn<#1|#2|#3>{{\langle#1|#2|#3\rangle}}
\def\vo(#1,#2)#3{{#1 \choose #2}_{\!#3}}
\def\zz#1.{\mathinner{(z-\zeta)^{#1}}}
\def\coupl(#1,#2){\big({#1\atop#2}\big)}
\def\F(#1,#2,#3,#4|#5|#6,#7,#8){\cF^{#2,#3}_{#1,#4}(#5|#6,#7,#8)}
\def\Fx(#1,#2,#3,#4|#5){\cF^{#2,#3}_{#1,#4}(#5)}

\makeatother
\begin{document}

\pubnumber{Imperial/TH/91-92/38 \\ hep-th/9209093}
\title{W-algebras arising as chiral algebras of conformal field
theory\thanks{
To appear in the proceedings of the XIX International Colloquium on
``Group Theoretical Methods in Physics'', Salamanca, Spain, June 29 --
July 4, 1992
}}
\author{
H. G. KAUSCH\thanks{Email: {\tt H\_G\_KAUSCH@V1.PH.IC.AC.UK}}
\address{
Theoretical Physics, Blackett Laboratory, Prince Consort Road, \\
Imperial College, London SW7~2BZ, U.K.
}}

\abstract{
It is argued that chiral algebras of conformal field theory possess a
W-algebra structure. A survey of explicitly known W-algebras
and their constructions is given.
}

\maketitle

\def\Wg{\mathord{\it Wg}}

\section{Introduction}

The presence of infinite dimensional {\em chiral} symmetries is
central to the study of
two dimensional conformal field theories.
These symmetries, called  {\em chiral algebras}, are formed by the
purely holomorphic and
purely anti-holomorphic fields, respectively.
How the field content of a conformal field theory (CFT) is organised
into representations of the chiral algebras is encoded in the fusion
ring.
The task of solving and classifying  general conformal field theories
can be split into
two separate problems: to find chiral algebras of conformal field
theory and to determine possible fusion
rings of chiral algebras.

This programme has been most successful for the Virasoro minimal
series \cite{BPZ}, where the fields fall into finitely many
representations of the Virasoro algebra.
This has subsequently enabled the complete characterisation of the
field content of such theories \cite{CIZu1}.
But considerable progress has also been made in constructing other
chiral algebras of CFT. I present here a survey of these results.

\section{Chiral algebras}

The (left) chiral algebra of a two dimensional conformal field theory
(CFT) is the symmetry algebra formed by all fields $\p(z)$ which
depend analytically on the coordinates.
Locality implies that
$[ \p(z), \psi(\z) ] = 0,$ for $z \neq \z$ and that all correlation
functions are meromorphic in the arguments with poles only at
coinciding arguments. From this one can derive the short distance
behaviour of chiral fields which is given by the
operator product expansion (OPE),
\begin{equation}
  \p(z) \psi(\z) = \sum_{n} \zz n. \c_n(\z),
\end{equation}
where the $\c_n(\z)$ are again local fields and only finitely many
singular terms appear.
Fields which
transform homogeneously under projective transformations
$
\g\colon z \mapsto (az+b)/(cz+d)
$
are called quasi-primary (or non-derivative).
They correspond to highest weights of the $\su(1,1)$ algebra generated
by the modes $L_0, L_{\pm1}$ of the stress tensor.
All other fields can be
obtained by taking derivatives. We will thus restrict attention to
quasi-primary fields and choose a basis $\{\chi^i\}$ with conformal
weights $\{ h_i \}$. Using $su(1,1)$-covariance and the usual contour
deformation argument we find from the OPE that
the commutator of two quasi-primary fields has the form
\begin{equation}
  {}[ \chi^i_m, \chi^j_n ] =
 \sum_k C^{ij}_k \, p^{h_ih_j}_{h_k}(m,n) \, \chi^k_{m+n}
+ D^{ij} {\textstyle {m+h_i-1 \choose 2h_i-1} \d_{m+n}}.
\end{equation}
The contributions from a particular $\su(1,1)$ representation are
collected in the $\su(1,1)$-Clebsch-Gordon coefficients
$p^{h_ih_j}_{h_k}(m,n)$ which are polynomials in $m$ and $n$ depending
only on the conformal weights of the fields involved
\cite{BFK+1}.
The $D^{ij} = \ip(\c^i,\c^j)$ define a metric on the space of
quasi-primary fields which can be used to raise and lower indices.
The field content of a particular chiral algebra is encoded in the
structure constants $D^{ij}$ and $C^{ijk} = \cfn<\c^i|\c^j(1)|\c^k>$.
Locality of the chiral algebra is equivalent to the Jacobi identity
for the commutator algebra which takes the form
\begin{equation}\label{eq:ass}
  \sum_{N=1}^{h_k+h_l-1}  \left(
	\sum_{q,q'} C^{ijq} C^{klq'} D_{qq'} \right) \,\,
	p^{h_jh_q}_{h_i}(m,n+p) p^{h_kh_l}_{h_q}(n,p)
	+ \hbox{cyclic} = 0.
\end{equation}
The study of chiral algebras is thus equivalent to the study of the
Lie algebra of quasi-primary fields.
However, we still have to deal with an infinite number of
quasi-primary fields.

In addition to the Lie bracket structure, which is determined solely
by the {\em singular} part of the OPE, chiral algebras admit a second
structure encoded in the {\em regular} part of the OPE, the so-called
normal ordered products (NOPs).
They are usually defined as the the constant term of the OPE,
$
  (\p\psi)(z) = \oint_z \,\, {d\z\over2\pi i} \,\,
	(\z-z)^{-1} \p(\z)\psi(z)
$.
However, $(\p\psi)(z)$ is not quasi-primary.  We therefore define a
new NOP, $\p\diamond\psi$, by projecting $(\p\psi)(z)$ onto
quasi-primary fields, i.e. $\p\diamond\psi$ is the new quasi-primary
field entering the OPE of $\p$ and $\psi$ at the constant term. The
$\diamond$-product is commutative, but still non-associative, and
differs from the usual NOP only by a finite number of correction terms
which are known explicitly \cite{BFK+1}.

We call fields which can be obtained from the $\diamond$-product {\em
composite}.
Fields which are orthogonal to all composite fields are called {\em
simple}.
The commutators of simple fields yields fields which are in general
not simple but involve NOPs of simple fields. The algebraic structure
given by commutators and NOPs of simple fields is called a {\em W-algebra}.

The simple fields generate the complete chiral algebra through NOPs and
derivatives and
all structure constants involving composite fields are determined by the
couplings between simple fields.
This means that the chiral algebra is completely determined by the
W-algebra.
Typically, a W-algebra
has only a finite number of simple fields reducing the problem of
constructing a chiral algebra to the {\em finite}
problem of finding solutions to the associativity constraints
(\ref{eq:ass}).

One starts be specifying the weights of simple fields, $(
h_1, \ldots, h_r)$ and then attempts to construct the W-algebra $W_{(
h_1, \ldots, h_r)}$ as follows.
\begin{itemize}
\item Construct all composite fields of weight less than twice the
maximum of the $h_i$.
These are the only fields which can appear in the commutator of two
simple fields.
\item Express the structure constants for composite fields in terms of
the structure constants for simple fields only.
\item Impose the Jacobi identity for simple fields.
\end{itemize}
This yields a system of polynomial equations for the simple structure
constants which one now has to solve. There are
three possibilities:
\begin{itemize}
\item $W_{( h_1, \ldots, h_r)}$ is consistent for generic values of $c$
({\em generic W-algebras}),
\item $W_{( h_1, \ldots, h_r)}$ is consistent for a finite set of
$c$-values ({\em exceptional W-algebras}),
\item $W_{( h_1, \ldots, h_r)}$ is inconsistent for all values of $c$.
\end{itemize}

This approach is systematic enough to be amenable to algebraic
computing.
The results obtained so far \cite{BFK+1,KWat1,Kaus1} fall into the
following classes.

\section{Generic W-algebras}

\paragraph{Current algebras:}
If the simple fields are all of weight one they form a current
algebra. Their commutator algebra is a Kac-Moody algebra and
the W-algebra is given by all derivatives and NOPs of
the currents.

\paragraph{Toda theories:}
W-algebras also appear as the chiral symmetry algebras of Toda field
theory. The W-algebra fields are normal ordered
differential polynomials of the Toda fields which commute with each of the
exponential terms in the Toda Lagrangian \cite{KWat3}.
The W-algebra $\Wg$ obtained from Toda field theory based on a Lie
algebra $g$ has $\rank g$ simple fields. Their weights are equal to
the orders of the independent Casimir operators of $g$ \cite{FFre6}. The
following quantum W-algebras have been constructed so far:
\begin{displaymath}
\begin{array}{cc}
\Wg & \hbox{weights} \\
\noalign{\vskip2pt} \hline \noalign{\vskip2pt}
\WA_1			& 2 \\
W(A_1\times U(1))	& 2,1 \\
W(A_1\times A_1)	& 2,2 \\
\WA_2			& 2,3 \\
\WC_2			& 2,4
\end{array}\qquad
\begin{array}{cc}
\Wg & \hbox{weights} \\
\noalign{\vskip2pt} \hline \noalign{\vskip2pt}
\WG_2	& 2,6 	\\
\WA_3	& 2,3,4	\\
\WB_3	& 2,4,6	\\
\WC_3	& 2,4,6	\\
W(A_1^n)& 2^n
\end{array}
\end{displaymath}
It has been realised recently that Toda field theories can be viewed
as Hamiltonian reductions of constrained WZNW theories \cite{FOR+2}.
In this context it is possible to associate (at least classically) a
W-algebra to every
integral $sl(2)$ embedding into a simple Lie algebra. In this case the
weights of the simple fields are one higher than the exponents of $g$
with respect to the $sl(2)$ embedding. However, no quantum version of
such a generalised algebra has been constructed as yet.

\paragraph{Orbifold models:}
Whenever a W-algebra has a (discrete) automorphism $\s$ the subspace of
fields invariant under $\s$ forms again a W-algebra. An example of
this is given by one of the algebras $W_{(2,4,6)}$, which turns out to
be the bosonic projection of the super-Virasoro algebra \cite{Bouw2}.

\section{Exceptional W-algebras}

\paragraph{Extensions:}
If for a specific $c$-value a generic W-algebra possesses
representations of integral weight with appropriate fusion
rules then it is possible to enlarge the generic W-algebra by these
representations to obtain an exceptional W-algebra.
This is indicated by the appearance of a non-diagonal modular
invariant \cite{BFK+1}. The simplest example is given by the
$(A_{p-1},D_{q/2+1})$ modular invariant of the Virasoro algebra. Here,
the field $\p_{p-1,q-1}$ has weight $h = (p-2)(q-2)/4$ and fusion rule
$\p_{p-1,q-1} \times \p_{p-1,q-1} = \id$. Whenever $p$ or $q$ is even,
this yields a W-algebra $W_{(2,h)}$. Exceptional
W-algebras  coming from such an extension of the Virasoro algebra are
listed below.
Extensions of other generic W-algebras have not yet been constructed.

\begin{displaymath}
\renewcommand{\arraystretch}{1.25}
\begin{array}{cl}
\hbox{weights} & \multicolumn{1}{c}{\hbox{$c$-values}} \\ \hline
2,2	& -{44\over5}, -{39\over10}, -{494\over35} \\
2,3	& -{114\over7}, -2, {4\over5} \\
2,4	& -{11\over14} \\
2,5	& -{350\over11}, -7, {6\over7} \\
2,6	& -{516\over13}, -{306\over55}
\end{array} \qquad
\begin{array}{cl}
\hbox{weights} & \multicolumn{1}{c}{\hbox{$c$-values}} \\ \hline
2,7	& -{25\over2} \\
2,8	& -{944\over17}, {21\over22}, -{224\over65} \\
2,3,3,3	& -2 \\
2,5,5,5	& -7 \\
2,7,7,7	& -{25\over2}
\end{array}
\end{displaymath}

\paragraph{Decoupling:}
It can happen that at specific values of $c$ some of the
simple fields of a generic W-algebra decouple and one is left with an
exceptional W-algebra.
The simplest example is the Ising-model $c=\half1$ which is in the
$\WE_8$ minimal series. In this case all simple fields apart from the
stress tensor decouple.
Known exceptional W-algebras arising from this decoupling procedure are:

\begin{displaymath}
\renewcommand{\arraystretch}{1.25}
\begin{array}{ccccl}
\Wg & \hbox{weights} & \longrightarrow &
\hbox{weights} & \multicolumn{1}{c}{\hbox{$c$-values}} \\ \hline
\WE_6 & 2,5,6,8,9,12 & \longrightarrow &
2,5 & -{350\over11}, -7, {6\over7} \\
\WE_8 & 2,8,12,14,18,20,24,30 & \longrightarrow &
2,8 & -{944\over17} \\
\WD_4 & 2,4,4,6 & \longrightarrow &
2,4,4 & 1, -{656\over11}
\end{array}
\end{displaymath}

These results support the
conjecture that all generic W-algebras
are all given by Hamiltonian reduction of constrained WZWN models and their
orbifolds.
The picture for exceptional W-algebras is much more complicated. While
the majority of cases fit the two procedures of extending and
decoupling there are a large number of cases which do not fall into
these patterns \cite{BFK+1,EFH+1}.

\newcommand{\noopsort}[1]{}

\end{document}